\documentclass[onecolumn, preprintnumbers, showpacs, nofootinbib, aps]{revtex4}
\usepackage{graphicx}

\newcommand{\bi}{\bibitem}
\newcommand{\be}{\begin{eqnarray}}
\newcommand{\ee}{\end{eqnarray}}

\begin{document}

\title{Accretion process onto super-spinning objects}

\author{Cosimo Bambi$^{\rm 1}$}
\email{cosimo.bambi@ipmu.jp}

\author{Katherine Freese$^{\rm 2}$}
\email{ktfreese@umich.edu}

\author{Tomohiro Harada$^{\rm 3}$}
\email{harada@rikkyo.ac.jp}

\author{Rohta Takahashi$^{\rm 4}$}
\email{rohta@riken.jp}

\author{Naoki Yoshida$^{\rm 1}$}
\email{naoki.yoshida@ipmu.jp}

\affiliation{$^{\rm 1}$Institute for the Physics and Mathematics 
of the Universe, The University of Tokyo, Kashiwa, Chiba 277-8568, Japan\\
$^{\rm 2}$The Michigan Center for Theoretical Physics, Department 
of Physics, University of Michigan, Ann Arbor, Michigan 48109, USA\\
$^{\rm 3}$Department of Physics, Rikkyo University, 
Toshima, Tokyo 171-8501, Japan\\
$^{\rm 4}$Cosmic Radiation Laboratory, the Institute of Physical and 
Chemical Research, Wako, Saitama 351-0198, Japan}

\date{\today}

\preprint{IPMU09-0113}

\begin{abstract}

The accretion process onto spinning objects in Kerr spacetimes 
is studied with numerical simulations. Our results show that 
accretion onto compact objects with Kerr parameter (characterizing 
the spin) $|a| < M$ and $|a| > M$ is very different. In the 
super-spinning case, for $|a|$ moderately larger than $M$, the 
accretion onto the central object is extremely suppressed due to a 
repulsive force at short distance. The accreting matter cannot reach 
the central object, but instead is accumulated around it, forming a 
high density cloud that continues to grow. The radiation emitted 
in the accretion process will be harder and more intense than the 
one coming from standard black holes; e.g. $\gamma$-rays could be 
produced as seen in some observations. Gravitational collapse 
of this cloud might even give rise to violent bursts. As $|a|$
increases, a larger amount of accreting matter reaches the 
central object and the growth of the cloud becomes less efficient. 
Our simulations find that a quasi-steady state of the accretion 
process exists for $|a|/M \gtrsim 1.4$, independently of the 
mass accretion rate at large radii. For such high values of 
the Kerr parameter, the accreting matter forms a thin disk at 
very small radii. We provide some analytical arguments to strengthen the
numerical results; in particular, we estimate the radius where 
the gravitational force changes from attractive to repulsive 
and the critical value $|a|/M \approx 1.4$ separating the two 
qualitatively different regimes of accretion. We briefly discuss 
the observational signatures which could be used to look for such 
exotic objects in the Galaxy and/or in the Universe.

\end{abstract}


\maketitle


\section{Introduction}

It is widely believed that the final product of the gravitational 
collapse of matter is a black hole (BH).  
In classical general relativity (GR), astrophysical BHs should 
be completely characterized by just three parameters: the mass 
$M$, the charge $Q$,  and the spin $J$. In this paper we focus 
on chargeless BHs.  The spin is often replaced by the Kerr parameter 
$a = J/M$. In classical GR, the values of $M$ and $a$ cannot be 
completely arbitrary, as they must satisfy the relation $|a| < M$, 
which is the condition for the existence of the horizon.  To see 
this we can examine the 3+1 dimensional Kerr solution. The 
position of the horizon is given by the expression~\cite{mtw,lppt}
\be
r_{H} = M + \sqrt{M^2 - a^2} \, .
\label{r-hor}
\ee 
It is clear that in (3+1)D spacetime the horizon cannot be formed if
\be
M < |a| \, .
\label{mass-limit}
\ee
In the absence of a horizon, there would be naked singularities
which are not allowed in GR. Indeed, if condition~(\ref{mass-limit}) 
is fulfilled, the Kerr metric makes it possible to reach the 
physical singularity at $r=0$ from some large $r$ in finite time 
without crossing any horizon. One could thus consider closed time-like 
curves and violate causality (see e.g. section~66c of~\cite{chandra} 
or ref.~\cite{carter}). For this reason, usually some kind of cosmic 
censorship is assumed and naked singularities are forbidden~\cite{penrose}. 
In particular, it is believed that naked singularities cannot be 
created by any physical process and therefore that the end-state of 
the gravitational collapse of matter is a Kerr BH with 
$|a| < M$~\cite{penrose}.

However, in this paper we consider objects which {\it do} violate the 
Kerr bound, i.e. with $|a|>M$. We call them ``super-spinars'', as
proposed in~\cite{horava}: since they have no event horizon, by the 
standard definition they are not BHs. Our main motivation is simple. The 
singularity 
can be viewed as the place where new physics should be expected: here 
observer-independent quantities like the scalar curvature diverge, 
while GR presumably breaks down above the Planck scale. It is therefore 
not unreasonable to expect that causality is conserved, not because 
the collapsing matter can form only objects with $|a| < M$, but because 
actually the central singularity is replaced by some high curvature 
region due to some quantum gravity effects, see e.g.~\cite{hn04,horava}.
In this case, there is apparently no reason to believe that the 
final product of the gravitational collapse of matter cannot have 
$|a| > M$.  Another possibility is that the collapsing matter forms a 
super-compact star with $|a| > M$ and exotic equations of state: now 
there is no central singularity, since the Kerr metric is a solution 
of Einstein equations only in vacuum; matter could have very exotic 
equation of state once it reaches densities so high that our knowledge 
of physics becomes inadequate. Actually, in general the metric at very 
small radii may deviate from the Kerr solution (the uniqueness theorem 
does not hold in absence of a regular horizon~\cite{robinson}), but in 
our discussion we will neglect such a possibility.

In this paper, we extend the studies started in~\cite{bf09, bft09}. 
The goal is to examine the main differences between the cases 
$|a| < M$ and $|a| > M$. Previous papers~\cite{bf09, bft09} discussed 
implications on the apparent shape. There it was found that, even 
if the bound is violated by a small amount, the shadow cast by the 
super-spinar (i.e. how it blocks light coming to us from an object behind it) 
changes significantly from the case with $|a| < M$: the shadow for 
the super-spinar is about an order of magnitude smaller as well 
as distorted.  This distinction can be used as an observational 
signature in the search for these objects.  Based on recent
observations at mm wavelength of the super-massive BH candidate at the 
Center of the Galaxy~\cite{doeleman}, the authors speculated on the 
possibility that it might violate the Kerr bound.

In this paper we discuss the process of accretion in a Kerr background
with arbitrary value of the Kerr parameter. For $|a|$ moderately larger 
than $M$, we find that the accreting matter cannot reach the central 
object, but is accumulated around it, forming a high density cloud. 
That may have interesting observational consequences. First, because 
of the high density and the high temperature of the plasma, the radiation 
produced in the accretion process can be much harder and more intense 
than the one coming from BHs. Second, there might be violent 
phenomena like bursts, when the amount of accumulated gas is large 
enough to gravitationally collapse. For higher values of $|a|$, the 
cloud evolves into a sort of disk, which is however very different
from the usual disk of accretion around a BH: here the disk
extends from $r \approx M$ to the center, leading to rapid accretion, 
increasing efficiently the mass of the central object, and producing 
hard radiation at very small radii. 

Unlike the case of Kerr BHs, we do not know if super-spinars are 
stable under small perturbations. Previous work has found that some 
very rapidly rotating objects in { 3+1} and higher dimensions 
can be unstable~\cite{emparan, cardoso}. To address this 
point regarding the super-spinars studied in this paper,
one should do a linearized analysis of the perturbations 
of these objects. However, the conclusion would be determined by 
the boundary conditions at the surface of super-spinars, which 
presumably depend on the quantum theory of gravity and are therefore 
unknown. Such a question cannot thus be addressed at present: here
we just assume that super-spinars are stable and we study the
accretion process onto these objects.

{\it Conventions}: We use natural units $G_N = c = k_B = 1$. The 
metric has signature $(-+++)$.

\section{Model and assumptions}

\subsection{Equations \label{sb-eqs}}

In this subsection we briefly review the basic ingredients of the 
formalism used in our study. For more details, see 
e.g.~\cite{banyuls, font, camenzind} and references therein.

We are going to simulate the accretion process of a test fluid
in a background curved spacetime; that is, we neglect the
back-reaction of the fluid to the geometry of the spacetime,
as well as the increase in mass and the variation in spin of the 
central object due to accretion. Such an approximation is surely
reasonable if we want to study a stellar mass compact object in
a binary system, because in this case the matter captured from
the stellar companion is typically small in comparison with
the total mass of the compact object. The results of our simulations
should not be applied to long-term accretion onto a super-massive 
object at the center of a galaxy, where accretion makes the mass 
of the compact object increase by a few orders of magnitude from 
its original value.

Our master formulas are the equations of conservation 
of baryon number and of the fluid energy-momentum tensor
\be
\label{eq-cons-j}
\nabla_\mu J^\mu &=& 0 \, , \\
\label{eq-cons-t}
\nabla_\mu T^{\mu\nu} &=& 0 \, ,
\ee
where $J^\mu$ and $T^{\mu\nu}$ are respectively the current of 
matter and the  fluid energy-momentum tensor
\be
J^\mu &=& \rho u^\mu \, , \\
T^{\mu\nu} &=& \rho h u^\mu u^\nu + p g^{\mu\nu} \, .
\ee
Here $\rho$ is the rest-mass energy density (for example, in the 
case of hydrogen plasma, $\rho = n(m_p + m_e)$, where $n$ is the 
number density of protons/electrons and $m_p$ ($m_e$) the proton 
(electron) mass), $p$ is the pressure, $u^\mu$ is the fluid
four-velocity, $h = 1 + \epsilon + p/\rho$ is the specific enthalpy, 
and $\epsilon$ is the specific internal energy density. In other
words, $\rho\epsilon$ is the thermal energy density ($\rho\epsilon=3 n T$ for 
non-relativistic hydrogen plasma), while $\rho(1+\epsilon)$ is the 
total energy density of the fluid. In order to solve the system, 
an equation of state $p = p(\rho,\epsilon)$ must be specified.

In the 3+1 formalism, the line element of the spacetime is written
in the form
\be\label{ds2}
ds^2 = - \left(\alpha^2 - \beta_i \beta^i\right) dt^2 
+ 2 \beta_i dt dx^i + \gamma_{ij} dx^i dx^j \, ,
\ee
where $\alpha$ is the lapse function, $\beta^i$ the shift 
vector and $\gamma_{ij}$ the 3-metric induced on each space-like slice. Greek indices
run from 0 to 3 and are lowered (uppered) by the 4-metric
$g_{\mu\nu}$ (inverse 4-metric $g^{\mu\nu}$). Latin
indices run from 1 to 3 and are lowered (uppered) by the 
3-metric $\gamma_{ij}$ (inverse 3-metric $\gamma^{ij}$).
For Eulerian observers, the 3-velocity of the fluid is given 
by
\be
v^i = \frac{u^i}{\alpha u^0} + \frac{\beta^i}{\alpha} \, ,
\ee
while the covariant components can be obtained by using
$\gamma_{ij}$, i.e. $v_i = \gamma_{ij} v^j$. For what follows,
it is convenient to use two sets of variables. The 
{\it primitive variables} are
\be
{\bf V} = \left( \rho, v^i, p \right)^T
\ee
and are the quantities whose evolution and quasi-steady state
(if any) we want to determine. The hydrodynamical equations are 
instead solved in term of the {\it conserved variables}
\be
{\bf U} = \left( D, S_i, \tau \right)^T \, ,
\ee
which can be written in term of the primitive ones as
\be
D &=& \rho W \, , \\
S_i &=& \rho h W^2 v_i \, , \\
\tau &=& \rho h W^2 - D - p \, ,
\ee
where $W = \alpha u^0 = (1 - v^2)^{-1/2}$ is the Lorentz factor
and $v^2 = v_i v^i$. The equations of conservation (\ref{eq-cons-j}) 
and (\ref{eq-cons-t}) can now be written as
\be\label{eq-cons-u}
\frac{1}{\sqrt{-g}} \left[
\frac{\partial}{\partial t} 
\left( \sqrt{\gamma} \, {\bf U} \right)
+ \frac{\partial}{\partial x^i} 
\left( \sqrt{-g} \, {\bf F}^i \right) 
\right] = {\bf \mathcal S} \, ,
\ee
where ${\bf F}^i$ and ${\bf \mathcal S}$ are defined by~\cite{banyuls}
\be
{\bf F}^i &=& \Big( 
D \left( v^i - \beta^i/\alpha \right) , \;
S_j \left(v^i - \beta^i/\alpha \right) + p \delta_j^i , \;
\tau \left(v^i - \beta^i/\alpha \right) + p v^i \; 
\Big)^T \, , \\
{\bf \mathcal S} &=& \Big( 
0 , \;
T^{\mu\nu} \left( \partial_\mu g_{\nu j} 
- \Gamma^{\lambda}_{\mu\nu} g_{\lambda j} \right) , \;
\alpha \left( T^{\mu 0} \partial_\mu \ln\alpha 
- T^{\mu\nu} \Gamma^{0}_{\mu\nu} \right) \; 
\Big)^T \, .
\ee
These equations are solved numerically by integrating over
the computational cells of the discretized spacetime.

\subsection{Background metric and test fluid}

In our study, the spacetime is described by the Kerr metric.
Using the Boyer-Lindquist coordinates, the line element~(\ref{ds2}) 
becomes
\be
ds^2 = - \alpha^2 dt^2
+ \frac{\Sigma^2 \sin^2 \theta}{\varrho^2} 
\left(d\phi - \omega dt \right)^2
+ \frac{\varrho^2}{\Delta} dr^2
+ \varrho^2 d\theta^2 \, , 
\ee
where $\alpha^2 = \varrho^2 \Delta / \Sigma^2$, 
$\omega = 2 a M r / \Sigma^2$, and $\Delta$, $\varrho^2$ and 
$\Sigma^2$ are defined by
\be
\label{eq:Delta}
\Delta &=& r^2 - 2 M r + a^2 \; , \\
\varrho^2 &=& r^2 + a^2 \cos^2 \theta \; , \\
\Sigma^2 &=& \left(r^2 + a^2\right)^2 - a^2 \Delta \sin^2 \theta \; . 
\ee

The equation of state of the accreting matter is the one
of an ideal gas with constant polytropic index $\Gamma$:
\be
p = \left(\Gamma - 1\right) \epsilon \rho \; .
\ee
In the simulations, we take $\Gamma = 5/3$ (non-relativistic gas).
The effects of electromagnetic fields on gas dynamics is neglected,
as well as non-adiabatic phenomena, like viscosity or heat
transfer.

\subsection{Calculation method \label{ss-calc}}

Our calculations are made with the relativistic hydrodynamics (RHD) 
module of the public available code PLUTO~\cite{pluto1, pluto2}, 
properly modified for the case of curved spacetime, as described 
in subsection~\ref{sb-eqs}. We do not solve the Riemann problem to 
compute fluxes, but we use a Lax-Friedrichs scheme; flux contributions 
are evaluated from all directions simultaneously (no dimensional
splitting); time evolution uses a second order Runge-Kutta algorithm.

The computational domain is the 2D axysimmetric space
$r_{in} < r < 20 \, M$ and $0 < \theta < \pi$, where $r_{in}$ 
is set according to the case under study. In this paper we present
the results for six different values of the Kerr parameter: 
$a = 0.4 \, M$, $a = 0.9 \, M$, $a = 1.1 \, M$, $a = 1.4 \, M$, 
$a = 2.0 \, M$, and $a = 3.0 \, M$. In the first two cases, we have 
a BH with event horizon at radial coordinate $r_H = 1.92 \, M$ and 
$r_H = 1.44 \, M$ respectively, and the inner boundary is set just 
outside the event horizon: $r_{in} = 2.00 \, M$ for $a = 0.4 \, M$ 
and $r_{in} = 1.50 \, M$ for $a = 0.9 \, M$. We adopt free-outflow 
boundary conditions, i.e. we set zero gradient across the inner boundary:. 
\be
\frac{\partial \rho}{\partial n} = \frac{\partial p}{\partial n}
= \frac{\partial v^i}{\partial n} = 0 \, ,
\ee
where $n$ is the coordinate orthogonal to the boundary plane.

For $|a| > M$, there is no event horizon but a naked singularity 
at $r = 0$. In the simulations presented in this paper, we took
$r_{in} = 0.5 \, M$ and we imposed no flow from the boundary. The 
choice of the value of $r_{in}$ can appear arbitrary. However, it
does not significantly alter the final result for any value of $|a|/M$, 
while for smaller $r_{in}$ the computational time increases 
considerably. As for the choice of the boundary condition, we 
have imposed no flow from the boundary, in order to prevent an 
unphysical injection of gas from the center.

Both for BHs and super-spinars, at the beginning of the 
simulations the density of the gas (plasma) around the compact objects is 
constant (and equal to $\rho_0$, see below) and we start injecting 
gas from the outer boundary at a constant rate. The gas is injected 
with radial velocity $v_r = - 0.0001$, while $v_\theta = v_\phi = 0$. 
We found that, for a given accretion rate, the final density profile 
is independent of the injection velocity for $v_r \lesssim - 0.001$. 
Indeed, for small values of the injection velocity, the evolution of 
the gas is determined by the gravitational force and not by the 
initial conditions.

The unit of length (and time) is the parameter $M$:
\be
M = 1.5 \cdot 10^6 \, 
\left(\frac{M}{10 \, M_\odot} \right) \, {\rm cm} \, , 
\ee
while the unit of density, $\rho_0$, is unspecified. The
mass accretion rate of the system in the simulations is
\be
\dot{M} = 3.4 \cdot 10^{17} \, 
\left(\frac{M}{10 \, M_\odot}\right)^2
\left(\frac{\rho_0}{10^{-3} \, {\rm g/cm^3}}\right) 
\, {\rm g/s} \, ,
\ee
or, in Eddington units\footnote{Since we assume that the process
is adiabatic, the unit of density $\rho_0$ can be left unspecified 
and our results can be applied (in principle) to any accretion 
rate, even orders of magnitude different. Such an assumption holds 
only for small accretion rates and optically thin gas. For example,
our results would predict no qualitative differences between 
sub-Eddington and super-Eddington accretions, which is definitively 
not true because for high accretion rates the pressure of 
radiation cannot be neglected.}, 
\be
\left( \frac{\dot M}{\dot{M}_E} \right) = 0.24 \,
\left( \frac{M}{10 \, M_\odot} \right)
\left( \frac{\rho_0}{10^{-3} \, {\rm g/cm^3}} \right) \, .
\ee
If we want to consider the accretion process onto an object
with a certain mass at a particular accretion rate, we 
have to adjust the unit of density $\rho_0$. 
For example, in the case of a super-massive object with
$M = 10^6 \, M_\odot$ and accreting at $10^{-6}$ its Eddington
limit, probably like the BH candidate at the Galactic Center,
one has to take $\rho_0 = 10^{-13} \, {\rm g/cm^3}$.

Because of our simple treatment of the accreting matter, the gas 
temperature becomes extremely high. Here we simply impose a maximum 
temperature: this work does not aspire to get an accurate
description of the accretion process, but only to figure
out the main differences between the accretion process onto
objects with $|a| < M$ and $|a| > M$. We have checked that the choice of 
$T_{max}$ does not change significantly the final results: for 
$T_{max} = 10 \, {\rm KeV}$, $100 \, {\rm KeV}$ and 
$1 \, {\rm MeV}$, the density changes by at most a 
few percent, while we do not expect that our simulations are 
as accurate.

\section{Results}

\subsection{Numerical study}

\begin{figure}
\par
\begin{center}
\includegraphics[height=6cm,angle=0]{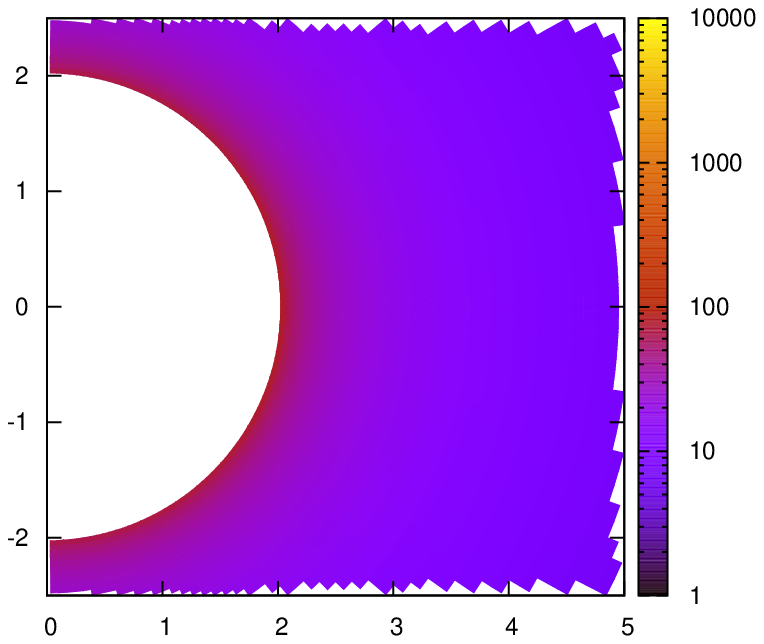} \hspace{.5cm}
\includegraphics[height=6cm,angle=0]{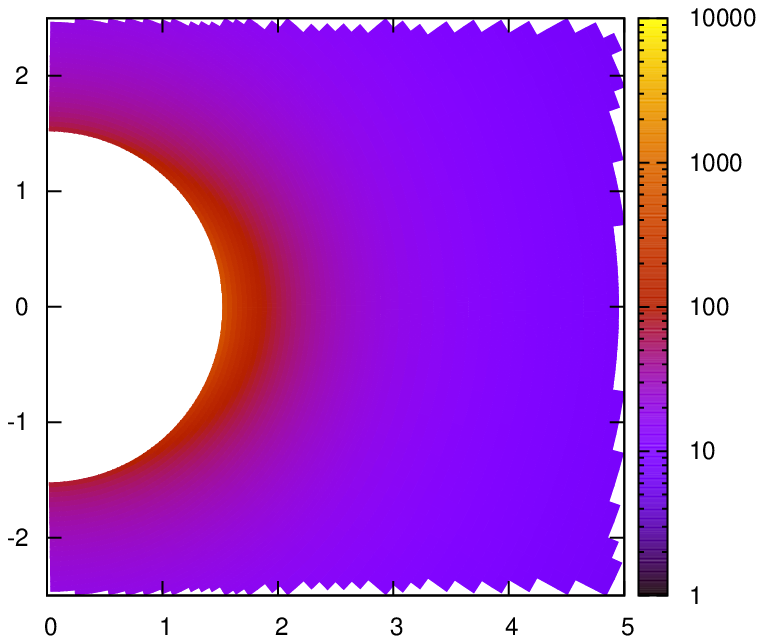} 
\end{center}
\par
\vspace{-5mm} 
\caption{Density of the accretion flow around a Kerr BH 
with $a = 0.4 \, M$ (left panel) and $a = 0.9 \, M$ (right panel).
The inner boundary is at $r_{in} = 2.00 \, M$ for the case 
$a = 0.4 \, M$, and at $r_{in} = 1.50 \, M$ for the case 
$a = 0.9 \, M$. The unit of length along the $x$ and $y$ axes 
is $M$. The density scale (shown on the right side of each figure)
ranges from $\rho_0$ to $10^4 \, \rho_0$, where $\rho_0$ is our
unit of density (see discussion in subsection~\ref{ss-calc}).}
\label{f-1}
\end{figure}

\begin{figure}
\par
\begin{center}
\includegraphics[height=6cm,angle=0]{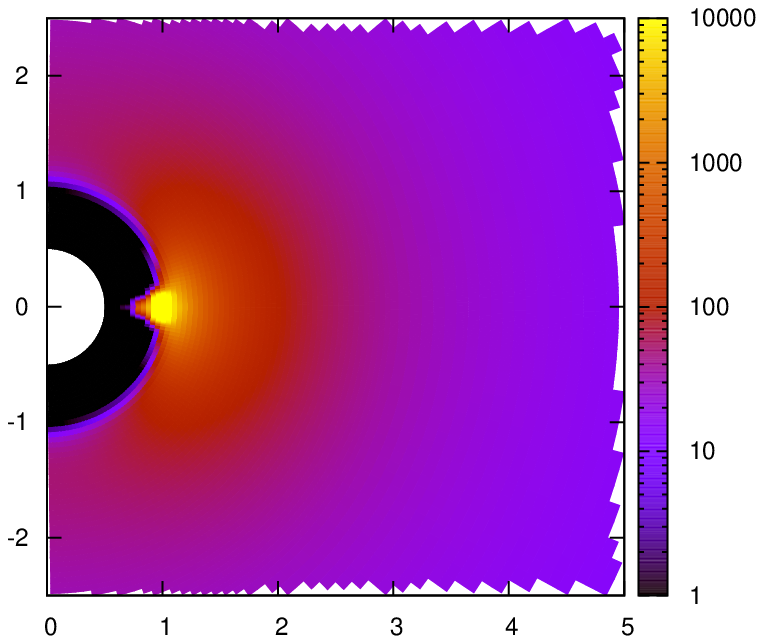} \hspace{.5cm}
\includegraphics[height=6cm,angle=0]{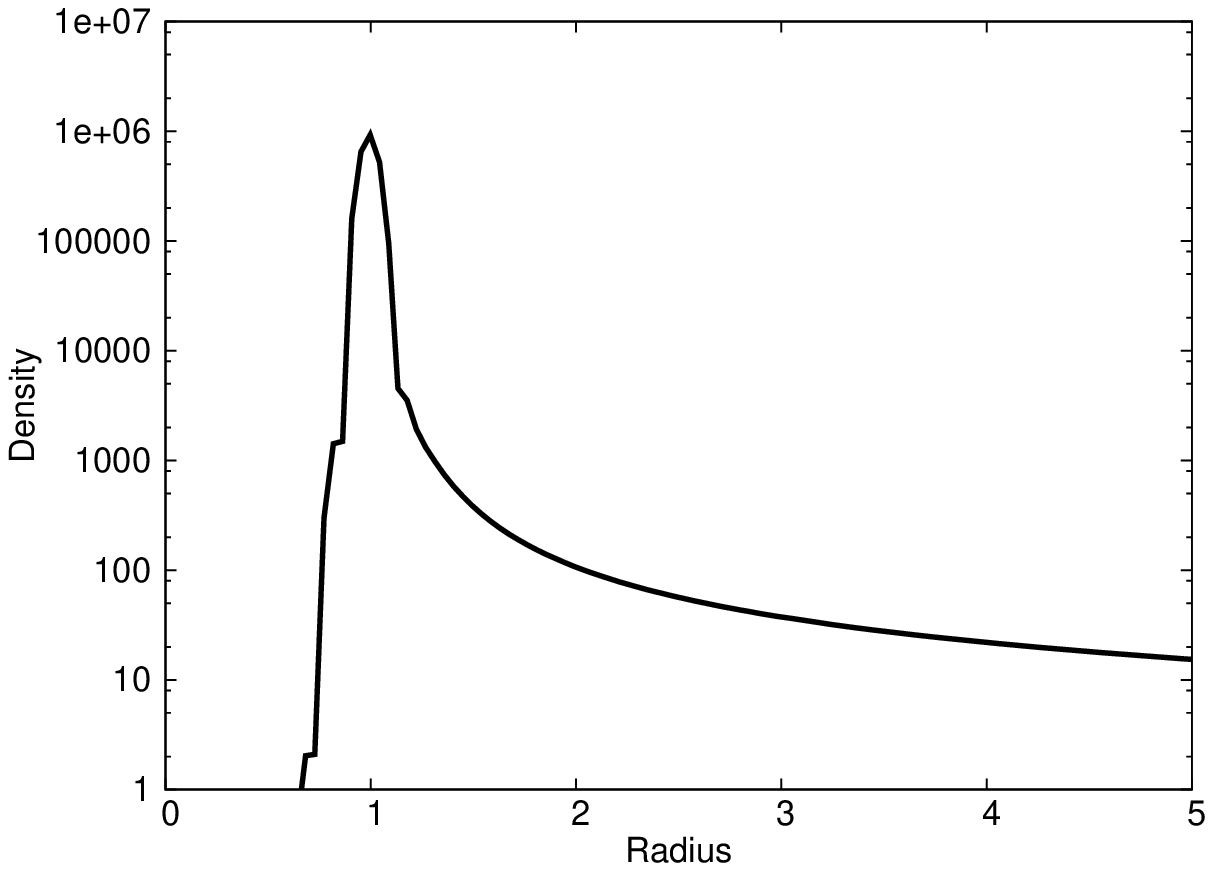} 
\end{center}
\par
\vspace{-5mm} 
\caption{Left panel: density  plot of the accretion flow 
around a super-spinar with $a = 1.1 \, M$ 
(color scheme as described in fig.~\ref{f-1}). Right panel: radial 
density profile on the equatorial plane of the same object. 
The inner boundary is at $r_{in} = 0.5 \, M$.}
\label{f-2}
\end{figure}

\begin{figure}
\par
\begin{center}
\includegraphics[height=6cm,angle=0]{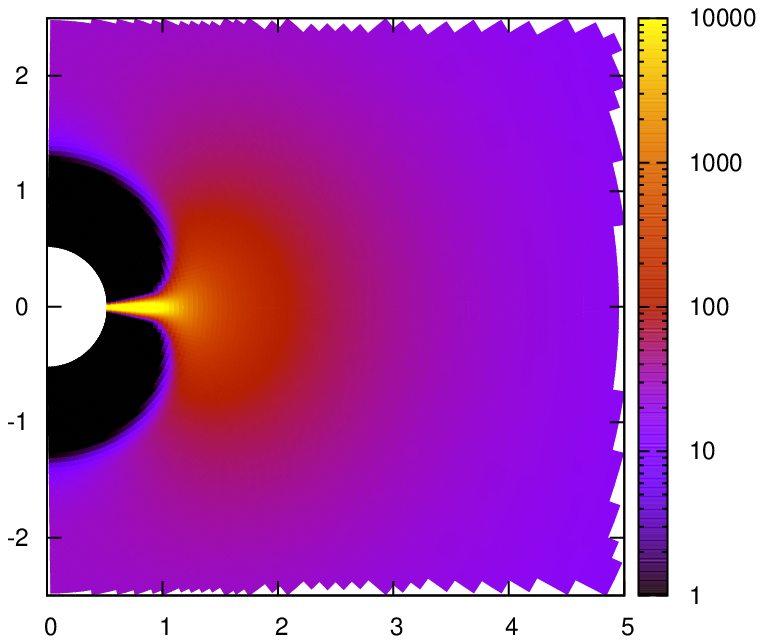} \hspace{.5cm}
\includegraphics[height=6cm,angle=0]{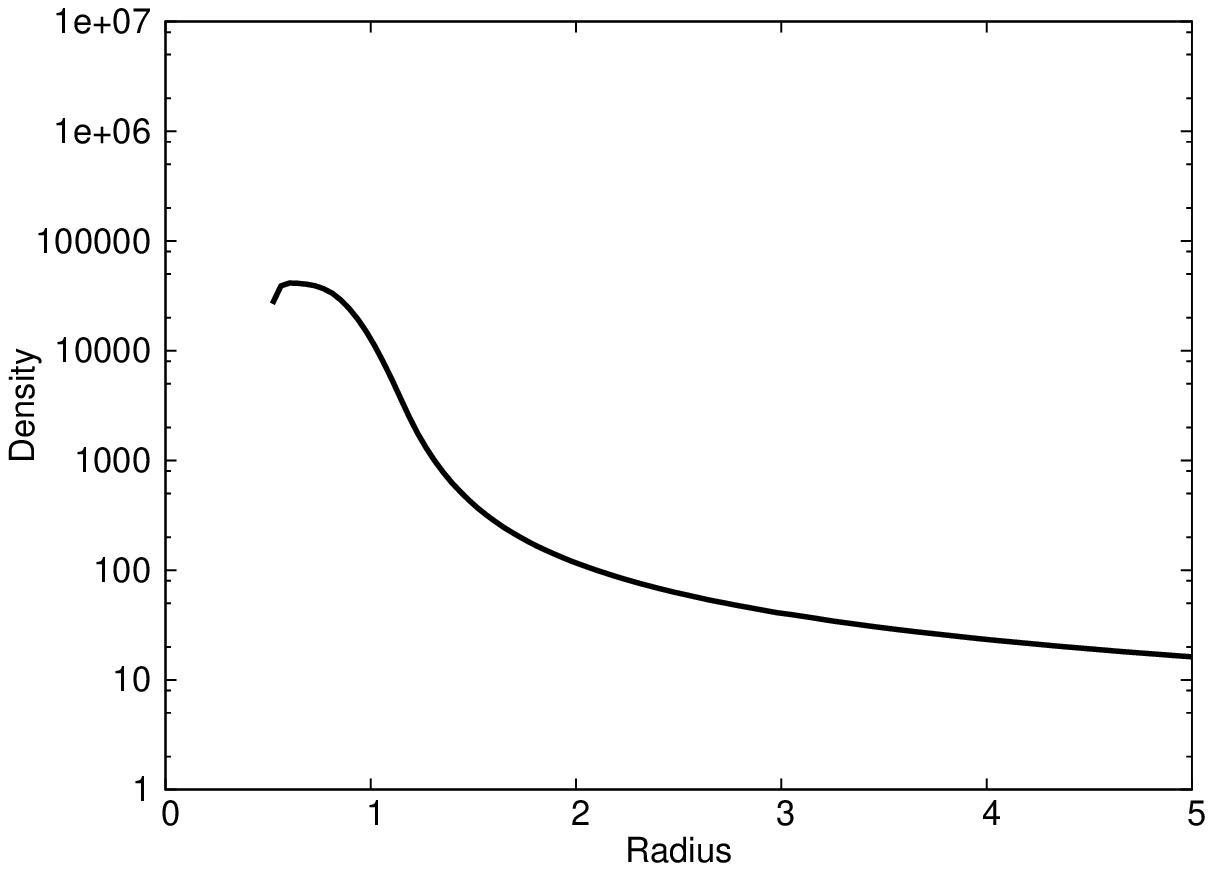} 
\end{center}
\par
\vspace{-5mm} 
\caption{Left panel: density plot of the accretion flow 
around a super-spinar with $a = 1.4 \, M$
(color scheme as described in fig.~\ref{f-1}). Right panel: 
radial density profile on the equatorial plane of the same 
object. The inner boundary is at $r_{in} = 0.5 \, M$.}
\label{f-3bis}
\end{figure}

\begin{figure}
\par
\begin{center}
\includegraphics[height=6cm,angle=0]{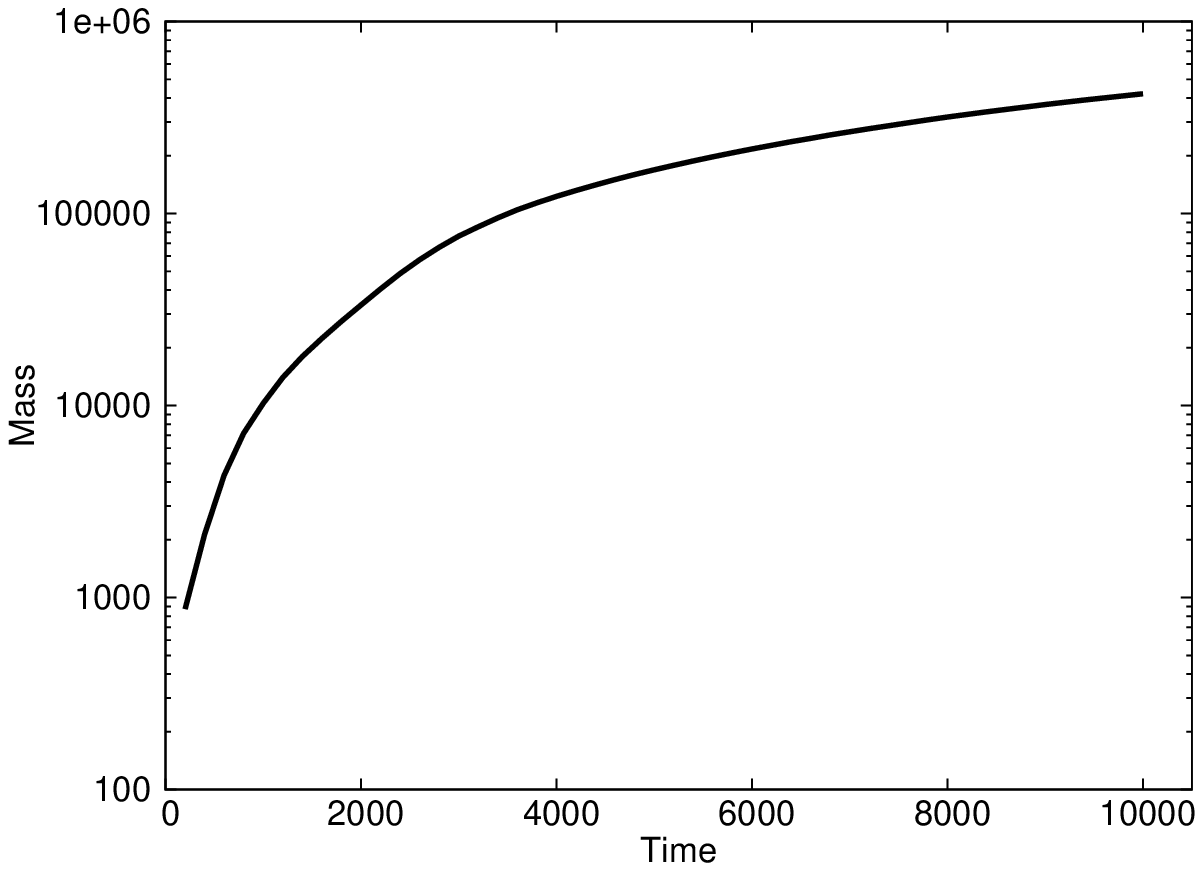} \hspace{.5cm}
\includegraphics[height=6cm,angle=0]{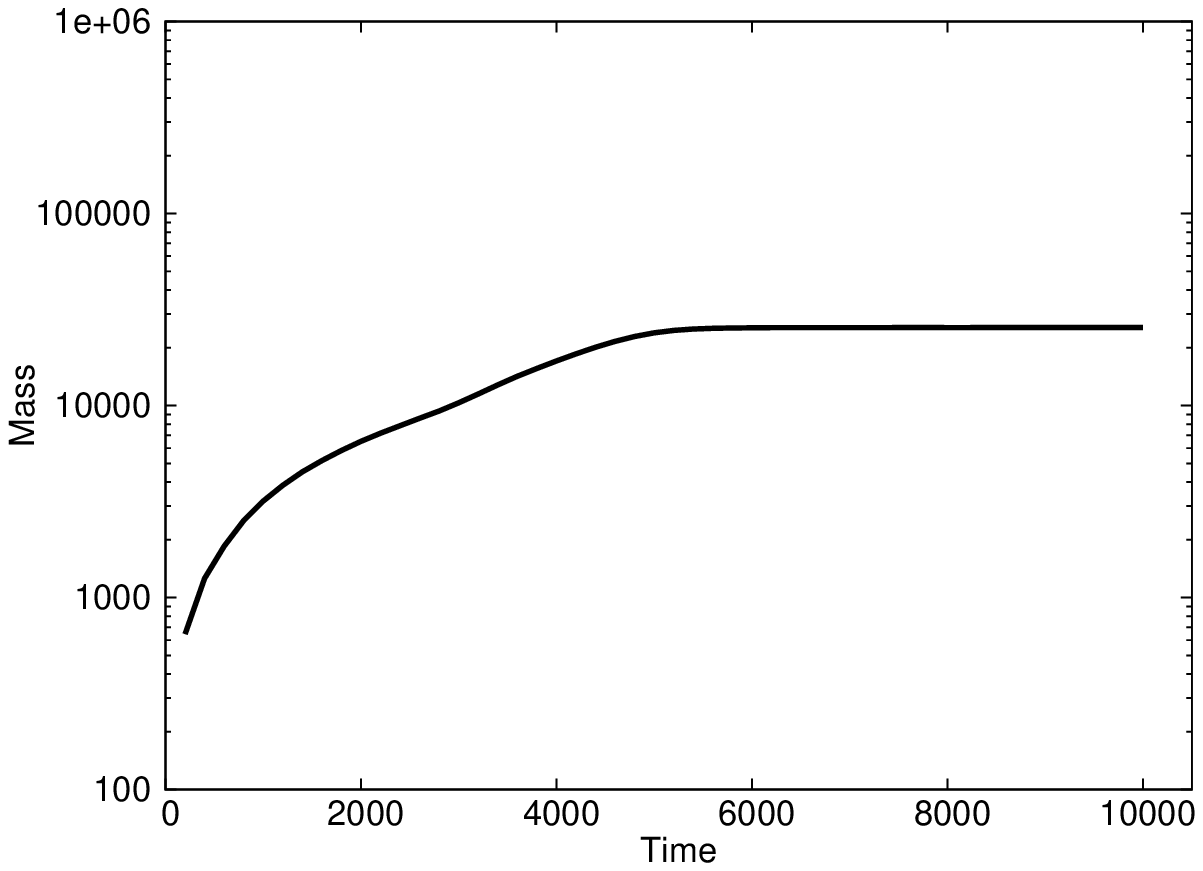} 
\end{center}
\par
\vspace{-5mm} 
\caption{Rest mass of the gas (in units $M^3 \rho_0$) as a
function of time (in units $M$) inside the region $0.5 \, M < r < 2.5 \, M$
in the cases of super-spinars with $a = 1.1 \, M$ (left panel)
and $a = 1.4 \, M$ (right panel). For values of the
Kerr parameter just above the Kerr bound, we do not find 
any quasi-steady state and the
gas is continuously accumulated around the object. For $a = 1.4 \, M$,
the flow reaches a quasi-steady state at $t \approx 6000 \, M$.}
\label{f-3}
\end{figure}

The results of the simulations are summarized in figs.~\ref{f-1},
\ref{f-2}, \ref{f-3bis}, \ref{f-3}, and \ref{f-4}. In fig.~\ref{f-1}, 
we present the density, $\rho$, of the accretion flow 
onto BH with Kerr parameter $a = 0.4 \, M$ (left panel) and 
$a = 0.9 \, M$ (right panel). The density scale in these and the 
other pictures ranges from $\rho_0$ to $10^4 \, \rho_0$, where 
$\rho_0$ is our unit of density, as discussed in subsection~\ref{ss-calc}.
The white area is out of the domain of computation. For both 
choices of $a$, the accretion flow reaches a quasi-steady state, 
that is, the density does not depend on time.

Fig.~\ref{f-2} (left panel) shows the density of the 
accretion flow around a super-spinar with $a = 1.1 \, M$. 
We use the same color intensity scale of 
fig.~\ref{f-1}, in order to facilitate the comparison between 
the cases $|a| < M$ and $|a| > M$. For the cells with density 
higher (lower) than $10^4 \, \rho_0$ ($\rho_0$), we use the 
same color scheme as though their density were $10^4 \, \rho_0$ 
($\rho_0$).
For example, a cell with $\rho = 10^6 \, \rho_0$ will have 
the same color as a cell with $\rho = 10^4 \, \rho_0$, and a cell
with $\rho_0 = 10^{-6} \, \rho_0$ will have the same color as
a cell with $\rho_0$.
As in the BH case, the white area is the region out of the 
domain of computation (now $r < 0.5 \, M$). In the right panel,
we present the radial density profile on the equatorial plane
(i.e. the plane with vertical axis $y = 0$ in the previous plot).

The peculiar feature of the case $a = 1.1 \, M$ is that the 
space around the central object is almost empty (the black 
region in the picture, where actually $\rho \ll \rho_0$). This 
is not the result of very high inflow velocity of the gas, but 
the effect of the repulsive force at short distances from the
center\footnote{Here the repulsive force is due to the singularity;
it is not the centrifugal force due to the rotation of the 
accreting gas. The fact that the gravitational force of a
time-like singularity might be repulsive in its neighborhood
has been already noticed in the literature, see e.g. 
refs.~\cite{kos,herrera}. The simplest example is the Schwarzschild 
spacetime with negative mass: there is no horizon, but a 
time-like singularity at $r = 0$, and the gravitational force 
is repulsive, as one can easily see by considering the 
Newtonian limit.}. The gas around the 
center is pushed away to larger radii, while the gas far from 
the object is attracted by the usual gravitational force. The 
object does not really accrete; instead matter accumulates around 
it, thus forming a high density cloud. Actually, on the 
equatorial plane there is some amount of gas falling into the 
super-spinar, but the fraction of matter with respect to the 
injected mass is so small that it is negligible. 
It is therefore clear that we cannot find any quasi-steady state 
in this case: matter is continuously accumulated into the cloud 
(fig.~\ref{f-2} shows the density profile at the time 
$t = 10^4 \, M$ of the simulation)\footnote{We notice that
a similar result was obtained in ref.~\cite{babichev} for the
case of Reissner-Nordstr\"om naked singularity: even there,
no quasi-steady accretion is possible and an ``atmosphere'' of 
fluid is formed around the center.}. Such a statement is checked 
by plotting the rest mass of the gas, defined by
\be
M_{gas} = \int \rho \sqrt{\gamma} \, d^3x \, ,
\ee
in the region $0.5 \, M < r < 2.5 \, M$, as a function of time 
(fig.~\ref{f-3}, left panel): as the time goes on, the total mass 
of the gas increases and there is no quasi-steady state.

As $|a|/M$ increases, more and more matter falls to the center and
leaves the domain of computation. The radial coordinate, say $r_{max}$,
of the point of the cloud with the highest density decreases (e.g.
$r_{max} \approx 1.0 \, M$ for $|a|/M = 1.1$, $r_{max} \approx 0.9 \, M$ 
for $|a|/M = 1.2$, and $r_{max} \approx 0.8 \, M$ for $|a|/M = 1.3$)
and the growth of the cloud slows down. For $|a|/M = 1.4$ and higher
values, the process changes significantly: $r_{max}$ approaches
$r_{in}$ and the process reaches a quasi-steady state; 
that is, the flux leaving the simulation at $r_{in}$ (moving towards 
the center) equals the flux of matter entering the simulation at 
$r_{out} = 20 \, M$. In fig.~\ref{f-3bis} we can 
see the density plot of a super-spinar with $a = 1.4 \, M$.
Still there is an empty region (black in the picture) in the
neighborhood of the center. However, that is not true near the
equatorial plane, where the density increases as $r$ decreases
and $r_{max} \approx r_{in}$, see the right panel of fig.~\ref{f-3bis}.
The right panel of fig.~\ref{f-3} shows that in this case the 
accreting flow reaches a quasi-steady state.

Since for $|a|/M \gtrsim 1.4$ the simulations show that $r_{max}$ 
goes to $r_{in}$, it is natural to wonder whether the threshold
$a = 1.4 \, M$ dividing two qualitatively different cases is
determined by the value of $r_{in}$. We have therefore run the
code with smaller values of $r_{in}$, finding however the same 
result: $r_{max} \to r_{in}$ and for $|a|/M \gtrsim 1.4$ the
process can reach a quasi-steady state. Our simulations thus
suggest that $|a|/M = 1.4$ is a critical value. 
We provide a qualitative explanation for the origin of
this particular value in the next subsection.

Fig.~\ref{f-4} shows the density plot for more rapidly spinning 
cases, with $a = 2.0 \, M$ (left panel) and $a = 3.0 \, M$ (right 
panel). The pictures show that the volume of the empty region around the 
compact object increases for higher values of the Kerr parameter. 
Yet the accreting matter reaches the inner boundary at $r_{in}$ 
more and more easily, thus leaving the computational domain and forming 
a structure similar to a disk. Such a disk is very different from 
the standard accretion disk around BHs. In very rapid 
super-spinars, the disk extends from $r \sim |a|/2$ to the 
center. At smaller radii, the disk becomes thinner and the density 
of the gas higher. On the other hand, the usual accretion disk
around BHs typically ranges from some very large radius, 
say $\sim 100 \, M$, inward 
to $r \sim r_{isco}$, where $r_{isco}$ is the innermost stable 
circular orbit (ISCO). The latter is $6 \, M$ for a Schwarzschild 
BH and as small as $M$ for a BH with $|a|/M = 1$. In addition to 
this, it is important to notice that here the disk appears in the 
case of initially spherically symmetric accretion, as the gas initially does
not have angular momentum. It is the result of the peculiar 
gravitational force at small radii around super-spinars.

\subsection{Qualitative Explanation}

The results of our simulations may seem too exotic and unexpected.
In this subsection, we present a few arguments to show that the 
emerging picture is physically reasonable.

First, let us see that the effective force can be repulsive at 
small radii. Assuming negligible velocity (i.e. 
$\dot{r} \approx \dot{\theta} \approx \dot{\phi} \approx 0$),
the radial component of the equations of motion for a massive 
test particle in Kerr spacetime is  
\be\label{eq-gr}
\ddot{r} \approx -\Gamma^r_{tt} \dot{t}^2
= - \frac{\Delta M \left(r^2 - a^2 \cos^2\theta\right)}{\varrho^6} 
\dot{t}^2 \, ,
\ee 
where 'dot' stands for a derivative with respect to the affine 
parameter and $\Delta$ is defined in Eq.(\ref{eq:Delta}).
 Since $\Delta > 0$ for any $r$ when $|a| > M$, the 
force is attractive or repulsive according to the sign 
of $(r^2 - a^2 \cos^2\theta)$. The same conclusion can be deduced 
from the master equations~(\ref{eq-cons-u}): if the fluid is 
non-relativistic, i.e. $v^2 \ll 1$ and $h \approx 1$, the radial 
component of eq.~(\ref{eq-cons-u}) becomes
\be\label{eq-fr}
\frac{\partial S_r}{\partial t} \approx
- \frac{\sqrt{-g}}{\sqrt{\gamma}} \Gamma^r_{tt} T^{tt} g_{rr}
= - \frac{\sqrt{-g}}{\sqrt{\gamma}} 
\frac{M \rho \left(r^2 - a^2 \cos^2 \theta\right) 
\Sigma^2}{\varrho^6 \Delta} \, ,
\ee
and the direction of the force still depends on the sign of 
$(r^2 - a^2 \cos^2\theta)$. In the case of Kerr BH with $|a| < M$, 
the force is always attractive because $(r^2 - a^2 \cos^2\theta) > 0$ 
for any $r > r_H > M$, while for $r < r_H$ the Kerr metric does 
not hold. We note also that, for the case of $r \gg |a|$ and 
$r \gg M$, eq.~(\ref{eq-fr}) reduces to the usual Newtonian case, 
with $- M \rho/r^2$ on the right hand side.

\begin{figure}
\par
\begin{center}
\includegraphics[height=6cm,angle=0]{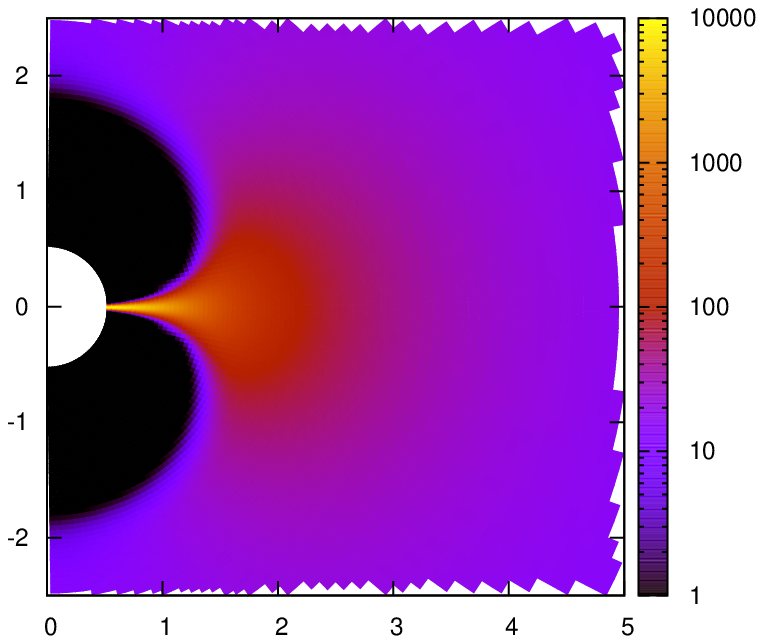} \hspace{.5cm}
\includegraphics[height=6cm,angle=0]{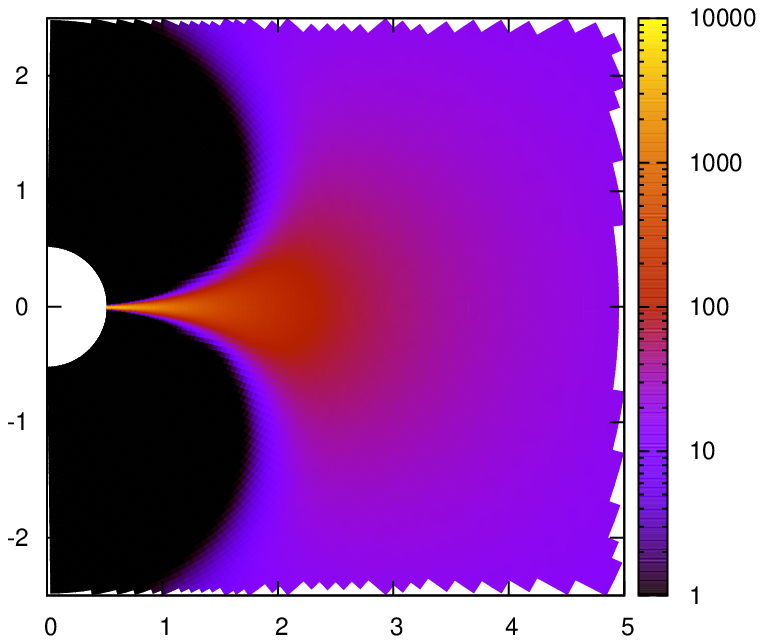} 
\end{center}
\par
\vspace{-5mm} 
\caption{Density plot of the accretion flow around a super-spinar
with $a = 2.0 \, M$ (left panel) and $a = 3.0 \, M$ (right panel).
Color scheme as described in fig.~\ref{f-1}.
The inner boundary is at $r_{in} = 0.5 \, M$.}
\label{f-4}
\end{figure}

Eqs.~(\ref{eq-gr}) and (\ref{eq-fr}) suggest that the force 
is repulsive inside the two spherical regions of radius $|a|/2$ and 
centers with coordinates $r = |a|/2$ and $\theta = 0$, $\pi$.
This is basically the shape and the size of the empty region around 
the center in the cases $a = 1.4 \, M$ (fig.~\ref{f-3bis}), $a=2.0\,M$, 
and $3.0\,M$ (fig.~\ref{f-4}), while it is less clear for 
$a = 1.1 \, M$. According to this 
interpretation, the origin of the repulsive force at small radii is 
a pure geometrical effect due to the spin $a$, and not the result 
of an effective force due to the angular momentum of the accreting 
plasma. We checked such a guess by performing some simulations 
injecting gas with different angular momenta. We found that the 
shape and the size of the empty regions around the center are 
essentially unaffected by the choice of the initial angular 
momentum of the gas, whose effect is instead to make the density 
profile asymmetric with respect to the equatorial plane.

The origin of the critical value $|a|/M \approx 1.4$ can be 
understood by slightly improving the previous argument. Since
inside the ergoregion everything must rotate, we relax the 
assumption $\dot{\phi} \approx 0$ and eq.~(\ref{eq-gr}) becomes
\be\label{eq-gr-2}
\ddot{r} \approx
- \Gamma^r_{tt} \dot{t}^2
- 2 \Gamma^r_{t \phi} \dot{t} \dot{\phi} 
- \Gamma^r_{\phi\phi} \dot{\phi}^2 \, ,
\ee
where
\be
\dot{t} &=& \frac{\Sigma^2 E - 2 a M r L_z}{\varrho^2 \Delta} \, , \\
\dot{\phi} &=& \frac{2 a M r E + (\varrho^2 - 2 M r) 
L_z \csc^2 \theta}{\varrho^2 \Delta} \, .
\ee
In order to study the sign of $\ddot{r}$, we need to specify the
constant of motion $E$ and $L_z$, which are respectively the energy
and the angular momentum along the spin (per unit mass) at infinity.
For a simple estimate, we can take $E=1$ (marginally bound orbits) 
and $L_z = 0$. Generally speaking, the first term on the right hand 
side of eq.~(\ref{eq-gr-2}) produces an attractive force at large 
radii and a repulsive force at small radii, while the second and 
third terms behave in the opposite way; that is, they give a force 
which is attractive and repulsive respectively for small and large 
$r$. It is easy to see that, on the equatorial plane, for $|a|/M$
slightly larger than 1, the force turns out to be attractive at
very small and large radii, and repulsive around $r = 0.75 \, M$.
The two values of $r$, say $r_1$ and $r_2$, for which the force is 
zero are given by the equation
\be\label{eq-acrit}
4 a^2 M r \Sigma^2 + 4 a^2 M \left(r^5 - a^2 M r^2\right) 
- \Sigma^4 = 0 \, ,
\ee   
where $\Sigma^2 = (r^2 + a^2)^2 - a^2 (r^2 - 2 M r + a^2)$.
Eq.~(\ref{eq-acrit}) reduces to the following simple equation
\be
a^4 + 2 a^2 r(r - 2M) + r^4 = 0 \, , 
\ee
which can be seen as a quadratic equation for $a^2$, with solution
\be\label{eq-a}
a^2 = r(2M-r) \pm 2 r \sqrt{M - r} \, .
\ee
As $|a|/M$ increases, the $r_1$ and $r_2$ converge to 
$r = 0.75 \, M$ and eventually coincide for
\be
a_{\rm crit}/M = \frac{3\sqrt{3}}{4} \approx 1.299 \, ,
\ee
which is the maximum of eq~(\ref{eq-a}). For larger values of the 
Kerr parameter, eq.~(\ref{eq-acrit}) has no solutions for $r>0$ and 
thus the force is everywhere attractive on the equatorial plane. 
Such a behavior explains the origin of the critical value 
separating the two different regimes of accretion found by our 
numerical study. The small discrepancy between the value obtained
in the simulations (about 1.4) and the one obtained
analytically (about 1.299) can be easily explained with
the difference between the motion of a fluid (for which the
hydrodynamical approximation holds, i.e. the mean free path of 
its particles is much smaller than the characteristic length 
scale of the system) and the one of a free particle. In particular,
the fluid occupies a finite volume and, even if for $|a|/M = 3\sqrt{3}/4$
the force is always attractive on the equatorial plane, that
is definitively not true out of the equatorial plane. The result
is that there is a sort of bottleneck effect at $r = 0.75 \, M$
and only when $|a|/M$ is slightly larger than $3\sqrt{3}/4$
all the falling gas can pass this critical point and the
process of accretion can reach a quasi-steady state.

\section{Astrophysical implications}

Our simulations show that the accretion process onto objects with 
$|a| < M$ and $|a| > M$ is very different. In the first case, 
the injected matter is swallowed by the object and the density 
profile reaches always a quasi-steady configuration: the BH 
grows at the same rate of the injection of matter. For super-spinars, 
accretion is more difficult due to a repulsive force
in the neighborhood of the center, except near the
equatorial plane. Here we can distinguish the cases $|a|/M < 1.4$ 
and $|a|/M \gtrsim 1.4$. For $|a|$ moderately larger than $M$, 
the repulsive force prevents that the accreting gas reaches
the central object: the mass of the super-spinar does not increase and
the gas is accumulated around the object. A cloud forms 
and grows; the matter density of this cloud becomes higher and 
higher. As shown in the left panel of fig.~\ref{f-3}, we do not 
find a quasi-steady state. As the quantity $|a|/M$ increases, 
more and more of the matter is able to fall all the way into the 
center. The highest density region of the cloud moves to smaller 
radii. Then, for $|a|/M \gtrsim 1.4$, the repulsive force in
the neighborhood of the center is no longer able to prevent a
regular accretion of the super-spinar. The accreting flow can reach a 
quasi-steady state (see the right panel of fig.~\ref{f-3}) by
forming a peculiar high density disk on the equatorial plane
at very small radii (figs.~\ref{f-3bis} and \ref{f-4}).

Limiting our considerations to a qualitative level, super-spinars
might explain observations like the ones reported in
ref.~\cite{pian}: the blazar PKS~0537-441 produces most of its 
flux in gamma rays, while for standard accretion onto usual Kerr BH 
it is not natural to produce gammas more than a few percent in 
bolometric luminosity~\cite{sbli}. Indeed, if the central object 
is a BH, one would expect an accretion luminosity proportional 
to $S T^4$, where $S \propto M^2$ is the emitting surface area and 
$T$ the temperature: a BH does not produce a significant
amount of hard radiation. In the 
case of super-spinars, the temperature of the accreting gas 
could become higher, because of much higher plasma densities, 
thus producing harder radiation.

It is however important to bear in mind that $i)$ some radiative
processes might become important and $ii)$ here the accreting matter 
is modeled as a test fluid. Actually, the particles of the accreting 
gas can lose energy and angular momentum through some dissipative
mechanisms, and that can have some effect on the accretion process, 
even if we expect that our basic conclusions still hold, especially
for small accretion rates. Secondly, the test fluid approximation breaks down when
the mass of the accreting gas is non-negligible in comparison
to the original mass of the object, e.g. that is true for 
super-massive objects at the center of galaxies. For $|a|/M < 1.4$, 
as the cloud grows, eventually its gravitational field becomes 
non-negligible, invalidating the test fluid approximation. Here 
it is difficult to predict what happens only from the present 
results, but we suggest a few possibilities. For example, the 
cloud could collapse onto the object, as soon as it wins against the 
repulsive force around the center. If the accretion rate at large 
radii is at 10\% of the Eddington limit, the mass of the cloud 
becomes comparable to the mass of the super-spinar after $10^8$~yr. The 
gravitational collapse of the cloud would be likely a violent event, 
and with two possible results: the formation of a BH with event horizon, 
i.e. an object with $|a| < M$, or the formation of a heavier 
super-spinar. The fate of the system presumably depends on the 
details of the accretion process and of the release of energy and 
angular momentum during the collapse of the cloud. Another possibility 
is that an event horizon just forms due to the accumulated matter, 
hides the mass from outside and no striking radiation is emitted. 
In the second scenario, the super-spinar would be the first 
product of the collapse, and then it would evolve into a Kerr BH, 
presumably without leaving any signature of its previous 
super-spinning stage.

\section{Conclusions}

In this paper we have studied the accretion process onto a spinning 
object. In particular, we have considered a Kerr spacetime with 
absolute value of the Kerr parameter $a$ (ratio of spin to mass) 
either smaller or larger than $M$, the mass of the object. In the 
first case, the spacetime contains a BH, in the second one a naked 
singularity. Our main motivation for considering the possibility of a
naked singularity is based on the observation that the singularity 
is more likely a pathology of classical GR and that in the full 
theory it must be replaced by something else. We do not know how 
the central singularity is resolved, but our results are probably 
not significantly affected by the details of the correct theory. 
The only relevant quantity astrophysically is likely to be $|a|/M$.

We can distinguish three cases:
\begin{enumerate}
\item BHs with $|a|/M<1$. We find the usual accretion picture: 
injected matter always reaches a quasi-steady state configuration, in 
which matter is lost behind the event horizon at the same rate as it 
enters into the computational domain. 
\item Super-spinars with $1<|a|/M<1.4$. Here the gas cannot reach 
the central object, because of a repulsive force in the neighborhood
of the center. As a result, the gas is efficiently accumulated around 
the super-spinar. That leads to the 
formation and growth of a high density cloud.
However, the accumulation process will stop at some point. One 
possibility is that it is interrupted by violent events due to the 
gravitational collapse of the cloud onto the object. This could be associated 
with the formation of a new object, either a BH or a heavier 
super-spinar. Another possibility is that the accumulated matter 
creates an event horizon, hiding the object from the outside, and with 
no abundant release of energy.
\item Super-spinars with $|a|/M \gtrsim 1.4$. Now the repulsive
force around the center is no longer capable of preventing a regular
accretion of the object. Our simulations find that the flow forms a high
density thin disk on the equatorial plane and reaches a quasi-steady 
state, i.e. matter enters and leaves the computational domain at the 
same rate. This disk is much closer to the center of the object than in the case 
of a standard BH. 
\end{enumerate}


\begin{acknowledgments}
We would like to thank Sergei Blinnikov, Hideki Ishihara,
Ken'ichi Nakao, and Lev Titarchuk for useful discussions at 
different stages of this work.
C.B. and N.Y. were supported by World Premier International 
Research Center Initiative (WPI Initiative), MEXT, Japan.
K.F. thanks the DOE and the MCTP at the University of Michigan 
for support.
T.H. was partly supported by the Grant-in-Aid for Scientific 
Research Fund of the Ministry of Education, Culture, Sports, 
Science and Technology, Japan (Young Scientists (B) 18740144 
and 21740190).
R.T. was supported by the Grant-in-Aid for Scientific Research 
Fund of the Ministry of Education, Culture, Sports, Science 
and Technology, Japan (Young Scientists (B) 18740144). 
\end{acknowledgments}


\end{document}